\PassOptionsToPackage{dvipsnames}{xcolor}
\documentclass[electronic]{vgtc}             % electronic version

\graphicspath{{figures/}{pictures/}{images/}{./}} % where to search for the images

\usepackage{times}                     % we use Times as the main font
\usepackage{tabu}                      % only used for the table example
\usepackage{booktabs}                  % only used for the table example
\usepackage{lipsum}                    % used to generate placeholder text
\usepackage{mwe}                       % used to generate placeholder figures

\usepackage{mathptmx}                  % use matching math font
\usepackage{siunitx}
\usepackage{graphicx}
\newcommand{\bpstart}[1]{\vspace{2pt}\noindent{\textbf{#1}}}

\newcommand{\mm}[1]{\SI{#1}{\milli\meter}}

\newcommand{\snd}[1]{\SI{#1}{\second}}

\newcommand{\met}[1]{\SI{#1}{\meter}}

\DeclareSIUnit[]
\pixel{pixels}

\usepackage[export]{adjustbox}
\newcommand{\lineIcon}{\includegraphics[height = 2.5ex, valign=c]{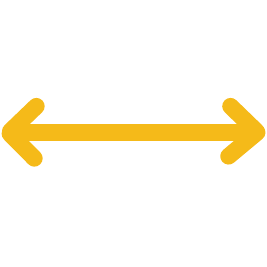}}
\newcommand{\circularIcon}{\includegraphics[height = 2.5ex, valign=c]{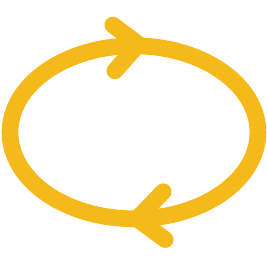}}
\newcommand{\infinityIcon}{\includegraphics[height = 2.5ex, valign=c]{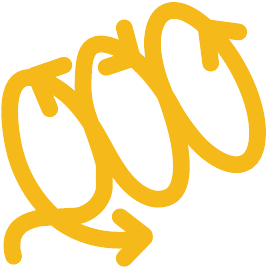}}
\newcommand{\slowWalkIcon}{\includegraphics[height = 2.5ex, valign=c]{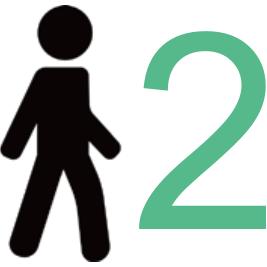}}
\newcommand{\fastWalkIcon}{\includegraphics[height = 2.5ex, valign=c]{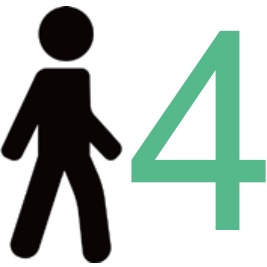}}
\newcommand{\slowJogIcon}{\includegraphics[height = 2.5ex, valign=c]{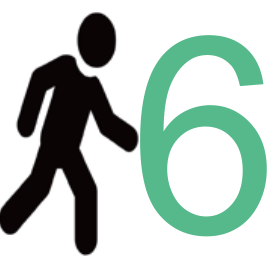}}
\newcommand{\standIcon}{\includegraphics[height = 2.5ex, valign=c]{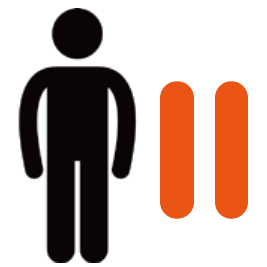}}
\newcommand{\sitIcon}{\includegraphics[height = 2.5ex, valign=c]{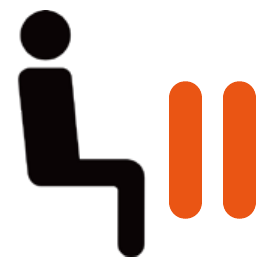}}

\usepackage{multirow}
\usepackage{multicol}
\newlength{\picturewidth}
\newlength{\pictureheight}
\newlength{\meanfigwidth}
\newlength{\difffigwidth}
\newlength{\meanfigheight}
\newlength{\difffigheight}

\usepackage{diagbox}
\usepackage{setspace}
\usepackage{calc}
\usepackage{fp}
\usepackage{textcomp}

\newlength\maxlen

\vgtcinsertpkg

\usepackage[export]{adjustbox}
\usepackage{xspace}
\usepackage{graphicx}
\usepackage{array}
\usepackage{subfig}
\usepackage{orcidlink}
\usepackage{microtype}

\usepackage[compact]{titlesec}

\titlespacing*{\section}
{0pt}{1.5ex plus 1.3ex minus .2ex}{0.5ex plus .2ex}
\titlespacing*{\subsection}
{0pt}{1ex plus 1.3ex minus .2ex}{0.5ex plus .2ex}

\onlineid{1159}

\vgtccategory{Evaluation}

\definecolor{isoblue}{HTML}{56B98B}
\definecolor{isoorange}{HTML}{EA5514}
\definecolor{isogreen}{HTML}{56B98B}
\definecolor{isoyellow}{HTML}{56B98B}
\definecolor{isot}{HTML}{F5BA1A}

\newcommand{\revision}[1]{\textcolor{black}{#1}}

\title{Micro Visualizations on a Smartwatch:\\Assessing Reading Performance While Walking}

\author{Fairouz Grioui\thanks{e-mail: fairouz.grioui@vis.uni-stuttgart.de \orcidlink{0009-0001-7358-6749}}\\ %
    \parbox{1.4in}{\scriptsize \centering University of Stuttgart Stuttgart, Germany} %
\and Tanja Blascheck\thanks{e-mail: research@blascheck.eu \orcidlink{0000-0003-4002-4499}}\\ %
    \parbox{1.4in}{\scriptsize \centering University of Stuttgart Stuttgart, Germany} %
\and Lijie Yao\thanks{e-mail: yaolijie0219@gmail.com \orcidlink{0000-0002-4208-5140}}\\ % 
    \parbox{1.4in}{\scriptsize \centering Universit{\'e} Paris-Saclay CNRS, Inria, LISN, France \\\scriptsize \& Xi'an Jiaotong-Liverpool University, China} %
\and Petra Isenberg\thanks{e-mail: petra.isenberg@inria.fr \orcidlink{0000-0002-2948-6417}}\\%
    \parbox{1.4in}{\scriptsize \centering Université Paris-Saclay CNRS, Inria, LISN, France}}

\abstract{
With two studies, we assess how different walking trajectories (straight line, circular, and infinity) and speeds (2\,km/h, 4\,km/h, and 6\,km/h) influence the accuracy and response time of participants reading micro visualizations on a smartwatch. We showed our participants common watch face micro visualizations including date, time, weather information, and four complications showing progress charts of fitness data. Our findings suggest that while walking trajectories did not significantly affect reading performance, overall walking activity, especially at high speeds, hurt reading accuracy and, to some extent, response time. Supplemental material is available at:~\url{https://osf.io/u78s6/}.} % end of abstract


\begin{document}


\firstsection{Introduction}

\maketitle
\revision{Smartwatches are wearable devices which people often use during activities that require them to be mobile. For example, smartwatches can keep athletes up-to-date about their performance, health, and well-being during a run. During leisurely walks, they can also provide other contextual information (e.\,g., weather and location data). Yet, it is unclear to which extent wearers moving around impacts reading micro visualizations of data on their smartwatch---and to which extent their movement characteristics, such as their speed or trajectory, matter. Answers to these questions are important because }previous studies conducted in the real world~\cite{Gouveia2018Invivo, pizza2016invivo} showed that, after checking the time, the main interactions with smartwatches consisted of inspecting notifications and using workout applications. Moreover, people engaged more with the device when physically active~\cite{Gouveia2018Invivo}. \revision{Given prior work on the challenges of reading smartphones while walking \cite{mustonen2004examining, schildbach2010investigating} we also expect challenges to exist for reading data from smartwatches on-the-go.}
Most of the previous lab studies exploring micro visualizations on smartwatches \revision{do not help to understand in which way performance may be impacted.} These studies were conducted in stationary settings while sitting on a chair and with a fixed device: the smartwatch was attached to a stand~\cite{blascheck2019glanceable, blascheck2023studies}, or worn on a non-moving arm~\cite{neshati2021sflg, neshati2019gsparks}. We complement this prior work by assessing to which extent a walking activity causing motion can impact the accuracy and response time of participants reading micro visualizations on a smartwatch. 
We conducted two studies to evaluate reading performance, task difficulty, and mental effort under three walking trajectories (Study~1), three walking speeds, and a sitting and standing baseline (Study~2). \revision{In general, we found that walking speeds had a greater impact on reading performance than walking trajectories. However, coordinating the reading task while following trajectory paths was more mentally demanding.}

\section{Related Work}
\revision{
We review prior work that explored micro visualization reading on mobile devices while sedentary and in motion.
}

\subsection{Micro Visualizations on Smartwatches}
Previous research in the visualization community contributed novel and space-efficient techniques to display complex micro visualizations, including large time series~\cite{chen2017visualizing}, multiple interlinked time series~\cite{neshati2021sflg}, and line charts with compressed axes~\cite{neshati2019gsparks}. Others have looked at people's ability to interpret simple micro visualizations (e.\,g., donut charts, bar charts~\cite{blascheck2019glanceable, blascheck2023studies, grioui2023heartrate}) and other conventional micro visualization techniques (e.\,g., hypnograms to represent sleep data~\cite{islam2022preferences}, and Time Spirals~\cite{suciu2018active} to self-track back pain) when scaled down to accommodate the small screen of a smartwatch. This past work has explored several avenues for using micro visualizations on small screens: using simple, low-dimensional visualizations~\cite{blascheck2019glanceable, blascheck2023studies}, creating techniques to adapt visualizations to the small screen~\cite{neshati2019gsparks}, or asking participants which visualizations they would prefer on the small screen~\cite{islam2022preferences}. \revision{Other prior work explored extending the display space to smartwatch straps \cite{Horak:2020:Smartwatch} for the display of data including line charts and maps.} However, most previous studies exploring micro visualizations on wrist-worn devices were conducted in stationary conditions. Reducing mobility when studying micro visualization reading on smartwatches helps control the viewing distance and angle variability. Nevertheless, it does not reflect realistic micro visualization reading conditions.

\subsection{Micro Visualization and Text Reading on-the-go}
Mobile devices %(e.\,g., typing or reading a text) 
have changed how individuals can consume and interact with information on-the-go. This has opened up new research to study dual-task performance~\cite{bruyneel2023texting, crowley2019effects, lin2017impact} to assess the interdependence between a primary engaging task and a secondary task. For instance, when a pedestrian has to handle multiple tasks simultaneously (e.\,g., reading a notification while walking), their cognitive abilities can be divided, causing one task to take precedence over the other. As a result, a decline in attention and performance can occur, increasing the probability of accidents~\cite{bruyneel2023texting, crowley2019effects, haga2015effects}. Prior work examined reading~\cite{mustonen2004examining, schildbach2010investigating} and texting~\cite{agostini2015texting, lin2017impact, nicolau2014mobileTextEntry} on a mobile phone while walking. The findings of these previous works suggest a decrease in performance, more workload, and a change in walking style, mainly in terms of gait speed reduction. Other studies investigated interacting with a smartwatch~\cite{fan2021adapting, neshati2021bezelglide, turner2018texting} while moving. By comparing walking and sedentary conditions, the authors found that dual-task walking affected the effectiveness of the interaction~\cite {fan2021adapting,turner2018texting}, as well as the walking style~\cite{fan2021adapting}. %---gait symmetry and step length 
Schiewe et al.~\cite{schiewe2020study} evaluated two micro visualizations and textual feedback on a smartwatch to guide runners doing forefoot training on a treadmill. The two visualizations showed both feet' heel and forefoot strikes as red and white horizontal bars, and participants received these visualizations well. %During this study, the treadmill was set to operate each participant's competitive running speed. 
However, the authors did not report on the effect of movement on reading the visualizations on the smartwatch.
\revision{We emphasize focusing on motion scenarios with relevant movement between viewers and micro visualizations. We do not discuss scenarios in which viewers and micro visualizations move together without relative movements (e.\,g., a car driver reading a fixed heads-up display \cite{Murtaza:2023:Autodrive}).}

\noindent In summary, research on the impact of motion on reading micro visualizations from a smartwatch is still limited, a gap we start to address with our current work. 
\section{Reading Micro Visualizations in Motion}

We conducted two studies to assess how motion affects the reading performance of viewers with micro visualizations while walking. We varied the walking trajectories in Study~1 and the walking speed in Study~2 to address the following research questions:

\bpstart{RQ1.} What is the influence of motion on the reading performance of micro visualizations on a smartwatch while walking?

\bpstart{RQ2.} What is the influence of different walking speeds and trajectories on the mental and physical load of the performed task?

\begin{figure}
     \centering
     \includegraphics[width=\linewidth]{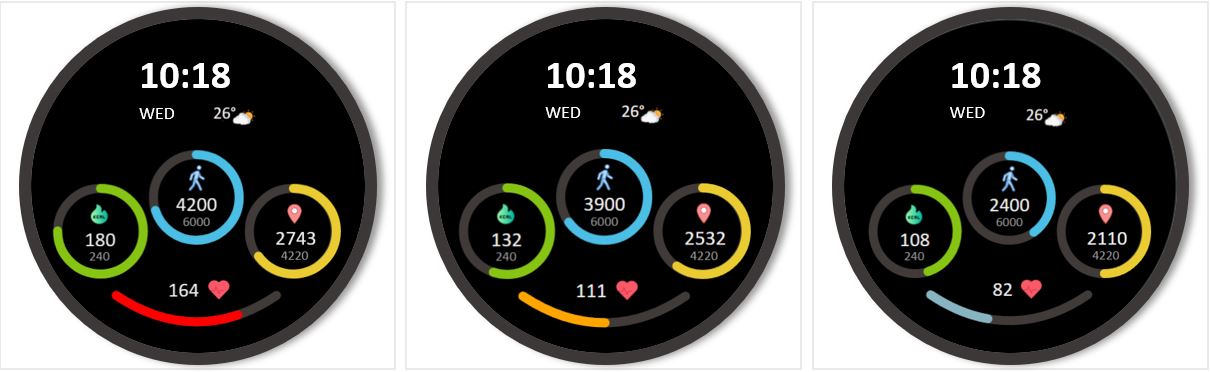}
    \caption{Examples of three watch-face stimuli showing the three radial progress charts: \textcolor{YellowGreen!50!ForestGreen}{\textbf{calories burned (cb)}}, \textcolor{Cyan!25!CornflowerBlue}{\textbf{step count (sc)}}, and \textcolor{yellow!75!Brown}{\textbf{distance walked (dw)}}. The visualized progress percentage on each stimulus are: \textbf{\textcolor{YellowGreen!50!ForestGreen}{cb: 75\%}}, dw: 70\%, sc: 65\% (left), \textbf{\textcolor{Cyan!25!CornflowerBlue}{sc: 65\%}}, dw: 60\%, cb: 55\% (middle), and \textbf{\textcolor{yellow!70!Brown}{dw: 50\%}}, cb: 45\%, sc: 40\% (right).}
    %\colorbox{darkgray}{backgroundcolor}
      \label{fig:stimuli}
      \vspace{-0.8em}
\end{figure}

\subsection{Task and Stimulus}
\label{sec:taskStimuli}
In this section, we describe our design choices for the displayed micro visualizations and the reading task. 

\subsubsection{Visualization Design Choices}
A watch face is the primary interface displayed on a smartwatch when activated. Typically, it shows about five complications\footnote{In horology, a complication is any graphical element presented on the watch face besides the time.} including a collection of glanceable information~\cite{islam2020visualizing}. We designed a watch-face stimulus showing the date, time, and weather information at the top and a set of fitness complications at the center and bottom of the screen (\autoref{fig:stimuli}). Rectilinear and radial progress charts (e.\,g., line and donut charts) are commonly used visual representations on watch faces \cite{blascheck2023studies} with radial representations being the most dominant. They often depict %battery level, humidity, and 
fitness data such as calories, distance, heart rate, and steps. Instead of showing exact quantities, radial representations are used to show progress toward a goal. Consequently, we chose radial and continuous progress charts to represent four fitness data. We displayed heart rate information %[50 -- 220 BPM] 
as a curved line progress chart. Its color depicts the Heart Rate Zone (HRZ) information (light blue: low, orange: moderate, red: high HRZ). At the center of the screen, we placed three radial progress charts with three distinct colors: \textcolor{YellowGreen!50!ForestGreen}{\textbf{calories burned}}, \textcolor{Cyan!25!CornflowerBlue}{\textbf{step count}}, and \textcolor{yellow!75!Brown}{\textbf{distance walked}}. The unfilled part of the bar was colored in gray. Inside each chart, we displayed %an icon in each radial progress chart corresponding to each complication, followed by 
text showing the current progress and target values. The values described a low to average activity level (i.\,e., 6,000 steps/day).

\subsubsection{Task Design}
Previous studies \cite{amini2017data, choe2017understanding} showed that people are interested in estimating the progress towards a goal when tracking their fitness data on-the-go. Therefore, our task asked: \emph{which fitness data has the highest progress towards a goal, and how much is this progress in percent?} Specifically, we asked participants to compare the proportions of the three fitness data radial charts, indicate the fitness data with the highest progress percentage, and estimate the percentage of the radial chart with the highest progress. We created nine distinct stimuli. Each three of them always showed one fitness data as the one with the highest progress of the three radial charts (\autoref{fig:stimuli}). 
All the proportion percentages we chose for our stimuli were multiples of~5. The difference between the radial progress chart with the highest progress and the two other radial progress charts was always 5\% and 10\%. We used the same stimuli in both studies, but we randomized their order between conditions. 

\subsection{Apparatus}
The studies were conducted in two different labs. The first had a walking space of 3\texttimes\met{5}. In the second, we used a Christopeit Sport TM2 Pro treadmill. We displayed the stimuli on a Fossil Gen 6 smartwatch with a screen size of \mm{44} and a resolution of 326 ppi.

\begin{table}[tb]
\begin{center}
        \caption{Participants' demographic data for the two studies.\label{tab:demographic}}
            \begin{tabular}{|m{.8cm}|m{1.8cm}|m{2.1cm}|m{1.9cm}|}
            \small \textbf{Study} & \small \textbf{Age in years}& \small \textbf{Height in cm}& \small \textbf{Arm-length in cm}\\
            \hline
            \small Study~1 &\small M$=$28.4, SD$=$4.5 &\small M$=$172.5, SD$=$9.4 &\small M$=$54.7, SD$=$4\\
            \small Study~2  &\small M$=$26.8, SD$=$4.7   &\small M$=$176.4, SD$=$10.7  &\small M$=$55.4, SD$=$4.6\\
            \end{tabular}
    \end{center}
    \vspace{-0.5cm}
\end{table}

%\subsection{Treadmill and Sedentary Conditions}
\begin{figure}[tb]
     \centering
      \includegraphics[width=1\linewidth]{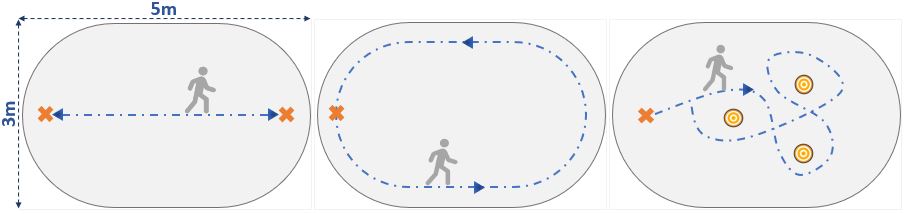}
      \caption{Illustrations of the 3 trajectories participants walked in Study~1, from left to right: {line, circular, and infinity trajectory}.}
      \label{fig:trajectories}
      \vspace{-0.8em}
\end{figure}

\subsection{Participants}
We recruited 18 participants:{ 4 \revision{women} and 14 \revision{men}} in Study~1, and  %age: 21-38 years (\emph{M}~$=$~28.4, \emph{SD}~$=$~4.5), height: 153-\cm{187} (\emph{M}~$=$~172.5, \emph{SD}~$=$~9.4), arm length: 45-\cm{61} (\emph{M}~$=$~54.7, \emph{SD}~$=$~4)}. 
24 participants:{ 9 \revision{women} and 15 \revision{men}} in Study~2; for details see \autoref{tab:demographic}. %, age: 19-37 years (M~$=$~26.8, SD~$=$~4.7), height: 155-\cm{195} (M~$=$~176.4, SD~$=$~10.7), arm length: 47-\cm{63} (M~$=$~55.4, SD~$=$~4.6)}. 
Eight participants from Study~1 and twelve from Study~2 reported that they owned a smartwatch or fitness band. Four and seven participants, respectively, from Studies~1 and 2 used the device on a daily basis. All participants were right-handed and apart from one participant in Study~1, \revision{all} reported wearing watches on their left wrist. \revision{All participants confirmed that they had good eyesight, including 10 in Study~1 and 8 in Study~2 who were wearing glasses.}

\subsection{Study Procedure and Design}
After signing the consent form, we explained the study task to the participants. Then, we tested their ability to read radial chart proportions, which all participants did correctly. In each study, participants performed all conditions and started with a training phase. The order of execution of the conditions was counterbalanced. \revision{Participants were shown 9 stimuli per condition, each for \snd{12} max}.

\subsection{Three Trajectories Conditions}
\label{sec:walkingtrajectories}
For Study~1, we designed three walking trajectories (\autoref{fig:trajectories}) to mimic different real-world walking situations.

\bpstart{\textbf{\lineIcon\ Line Trajectory.}} One-way straight trajectory, with known start and end points. Participants took nine laps, one lap per stimulus. 

\bpstart{\textbf{\circularIcon\ Circular Trajectory.}} A continuous and oval-shaped trajectory.

\bpstart{\textbf{\infinityIcon\ Infinity Trajectory.}} Participants walked continuously around three cones, creating an infinity-like trajectory.%\newline

\vspace{2mm}During the \lineIcon\ line trajectory condition, the walked distance was quite short. Therefore, we asked participants to press the smartwatch screen at the beginning of each lap to trigger the display of the visualization and ensure that participants read the micro visualizations while walking. In contrast, for both the \circularIcon\ circular and the \infinityIcon\ infinity trajectories, participants pressed the screen of the smartwatch only once at the beginning when they started walking. Every \snd{12}, a new visualization was displayed on the screen, and participants were notified with a vibration from the smartwatch to look at the stimulus. We did not prompt participants to perform the task as fast as possible. However, they were informed about the \snd{12} delay.

\subsection{Walking Speed and Sedentary Conditions}
In Study~2, we tested the following three speeds on the treadmill.

\bpstart{\textbf{ \slowWalkIcon\ Slow Walking Pace.}} In the first study, participants walked at a speed of 2.5\,km/h on average. Therefore, we set the slow walking speed on the treadmill to 2\,km/h to simulate a slow walking pace. 

\bpstart{\textbf{\fastWalkIcon\ Fast Walking Pace.}} We operated the treadmill with a speed of 4\,km/h, to simulate a fast walking pace.

\bpstart{\textbf{\slowJogIcon\ Slow Jogging Pace.}} We set a jogging speed of 6\,km/h to investigate a moderately accelerated condition.

\bpstart{\textbf{Sedentary Conditions.}} Participants also performed the reading task while \sitIcon\ sitting on a regular chair and while \standIcon\ standing. 
\subsection{Measurements}

We recorded participants' spoken answers on video and in a separate spreadsheet to avoid interrupting their movement. In addition, we collected response time and subjective measurements.

\bpstart{\textbf{Reading Accuracy.}} We analyzed the correctness of the answers about the fitness data with the highest progress (fitness data comparison accuracy). For all correct answers, we analyzed the difference between the provided and the correct percentages (proportion estimation accuracy). 

\bpstart{\textbf{Response Time.}} We measured the time participants took to verbalize their response for each stimulus when their arm was in a stable reading position. We extracted this information with a video annotation tool. In Study~2, we only analyzed 70\% of the videos because some of the video records were missing some or all trials.

\bpstart{\textbf{Subjective Measurements.}}
For each condition, participants used a 7-point Likert scale to rate the task difficulty (1:~not at all difficult, 7:~very difficult), their confidence with their answers (1:~not at all confident, 7:~very confident), and the mental and physical load of the task (1:~not at all demanding, 7:~very demanding).
\section{Results and Discussion}
We report sample means of the reading performance for 486 trials from Study~1 and 1080 trials from Study~2 using 95\% confidence interval (CI) estimation~\cite{besancon2019continued,cockburn2020threats,cumming2011understanding,dragicevic2016fair} using BCa bootstrapping (10,000 bootstrap iterations). In addition, we report the differences between means to compare the three conditions. We adjusted the CIs of mean differences for multiple comparisons using Bonferroni corrections~\cite{higgins2004introduction} to ensure consistency between the two studies. When reading a CI of mean differences, a non-overlap of the CIs with 0 is evidence of a difference. In this section, we reflect on the key findings from the results of the two studies with respect to our research questions and present takeaway messages. 

\subsection{RQ1: The Influence of Motion on Reading Performance} We compare reading performance within the two studies' walking conditions, as well as between walking and sedentary conditions in Study~2 to assess the influence of multitasking and movement on reading micro visualizations.

% measurements Study~1
\setlength{\picturewidth}{.45\linewidth}
\setlength{\pictureheight}{1.65cm}
\setlength{\meanfigwidth}{.1\linewidth}
\setlength{\meanfigheight}{1.3cm}
\setlength{\difffigwidth}{.2\linewidth}
\setlength{\difffigheight}{1.4cm}
\begin{table}[tb]
\renewcommand{\arraystretch}{1.9}   
\centering
    \caption{Means of the reading performance measurements for Study~1: response time (Row 1), fitness data comparison accuracy (Row 2), proportion estimation accuracy (Row 3). Error bars represent 95\% Bootstrap confidence intervals (CIs).% in black, adjusted for three pairwise comparisons with Bonferroni correction (in red).
    \\
    \small\lineIcon\ line trajectory, \circularIcon\ circular trajectory, \infinityIcon\ infinity trajectory
    \vspace{-0.3cm}\label{tab:study01_measurements}}
    \scriptsize
    \centering
    \fontsize{5.1}{5.3}\selectfont
    \begin{tabular}
    {@{}p{1.4\picturewidth}@{}%for the 1st image
    @{}r@{~}%for writing the average
    @{}p{1cm}@{}%for writing the CI
    % @{}p{1.8\picturewidth}@{}%for the 2nd image
    % @{}r@{~}%for the average
    % @{}p{0.8cm}@{}%for writing the CI
    }
    %%%%%%%% Title row: response time %%%%%%
    \specialrule{0em}{0ex}{0ex}
    \multicolumn{1}{c}{\small \textbf{Response Time -- in seconds: Study~1}} \\
   \specialrule{0em}{0ex}{2ex}
    \multirow{1}{*}{\includegraphics[height=\meanfigheight]{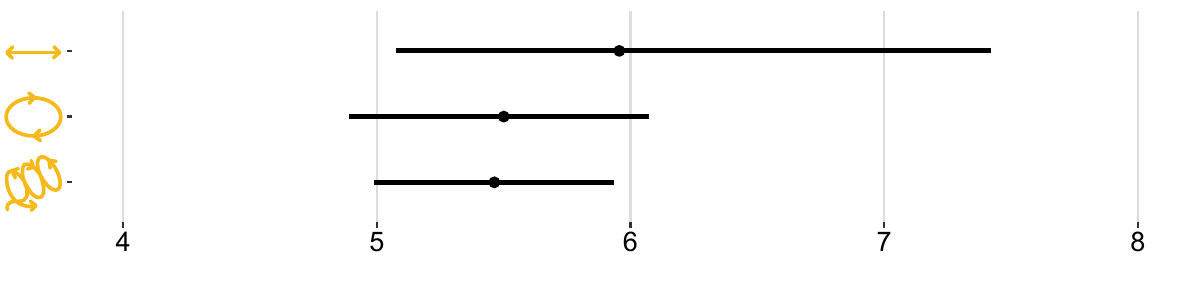}}&%Image: Alpha mean
    5.96,&% mean alpha - line
    [5.07,7.42]% CI alpha - line
    \\
    %% row 1.2 
    &%image row stays empty
    5.50&% mean alpha - circular
    [4.89,6.07]% CI alpha - circular
    %image row stays empty
    \\
    %% row 1.3
    &%image row stays empty
    5.46&% mean alpha - cones
    [4.99,5.93]% CI alpha - cones
    %image row stays empty
    \\ 
    \specialrule{0em}{0ex}{2ex}
    %% subfigures title
    \multicolumn{1}{c}{\fontsize{7}{7}\selectfont Means with CIs}\\ 
%    \specialrule{0em}{0ex}{2ex}
    % \multirow{1}{*}{\includegraphics[height=\meanfigheight]{figures/results/study1/mean_response-time-diff.pdf}}&%Image: Beta mean
    % -0.04&% mean beta - line
    % [-0.43,0.30]% CI beta - line
    % \\
    % &
    % -0.49&% mean beta - circular
    % [-1.79,0.31]% CI beta - circular
    % \\
    % &
    % 0.46&% mean beta - cones
    % [-0.30,1.86]% CI beta - cones
    % \\ 
    % \specialrule{0em}{0ex}{2ex}
    % \multicolumn{1}{c}{\fontsize{7}{7}\selectfont Pairwise differences with CIs}\\
    %%%%%%%% Title row: response time %%%%%%
    \cline{1-3}
    %%%%%%%% Title row: fitness data Accuracy %%%%%%
    \specialrule{0em}{0ex}{2ex}
    \multicolumn{1}{c}{\small \textbf{Fitness Data Comparison Accuracy -- in percent: Study~1}} \\
    \specialrule{0em}{0ex}{2ex}
    \multirow{1}{*}{\includegraphics[height=\meanfigheight]{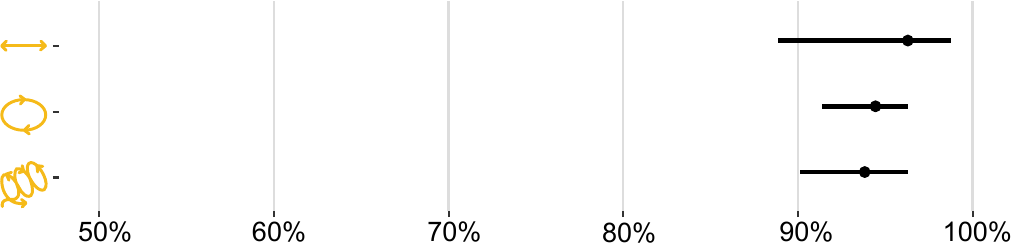}}&%Image: Alpha mean 
    96.3&% mean alpha - line
    [88.9,98.8]\\% CI alpha - line
   
    %% row 1.2 
    &%image row stays empty
    94.4&% mean alpha - circular
    [91.4,96.3]\\% CI alpha - circular
   
    %% row 1.3
    &%image row stays empty
    93.8&% mean alpha - cones
    [90.1,96.3]\\% CI alpha - cones
    \specialrule{0em}{0ex}{4ex}
    %% subfigures title
    \multicolumn{1}{c}{\fontsize{7}{7}\selectfont Means with CIs}\\
    % \specialrule{0em}{0ex}{2ex}
    
    % \multirow{1}{*}{\includegraphics[height=\meanfigheight]{figures/results/study1/means-correct-data-diff.pdf}}&%Image: Beta mean
    % -0.6&% mean beta - line
    % [-5.6,4.3]% CI beta - line
    % \\
    % &%image row stays empty
    % -2.5&% mean beta - circular
    % [-8.6,4.9]% CI beta - circular
    % \\
    % &%image row stays empty
    % 1.9&% mean beta - cones
    % [-4.3,6.2]% CI beta - cones
    % \\
    % \specialrule{0em}{0ex}{2ex}
    % %% subfigures title
    % \multicolumn{1}{c}{\fontsize{7}{7}\selectfont Pairwise differences with CIs} \\
    % \specialrule{0em}{0ex}{2ex}
    %%%%%%%% Title row: Fitness Data Accuracy %%%%%%
    \cline{1-3}
    %%%%%%%% Title row: Answer Accuracy %%%%%%
    \specialrule{0em}{0ex}{2ex}
    \multicolumn{1}{c}{\small \textbf{Proportion Estimation Accuracy -- in percent: Study~1}} \\
    \specialrule{0em}{0ex}{2ex}
    \multirow{1}{*}{\includegraphics[height=\meanfigheight]{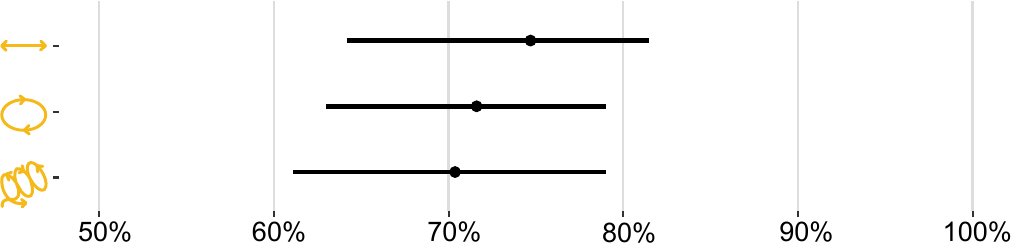}}%Image: Alpha mean
    &74.7% mean alpha - line
    &[64.2,81.5]% CI alpha - line
    \\
    %% row 1.2 
    &%image row stays empty
    71.6&% mean alpha - circular
    [63.0,79.0]% CI alpha - circular
    \\
    %% row 1.3
    &%image row stays empty
    70.4&% mean alpha - cones
    [61.1,79.0]% CI alpha - cones
    \\
    \specialrule{0em}{0ex}{4ex}
    %% subfigures title
    \multicolumn{1}{c}{\fontsize{7}{7}\selectfont Means with CIs}
    \\
    % \specialrule{0em}{0ex}{2ex}
    % \multirow{1}{*}{\includegraphics[height=\meanfigheight]{figures/results/study1/mean-correct-percentage-diff.pdf}}&%Image: Beta mean
    % -1.2,&% mean beta - line
    % [-11.1,8.5]% CI beta - line
    % \\
    % &%image row stays empty
    % -4.3&% mean beta - circular
    % [-16.0,8.0]% CI beta - circular
    % \\
    % &%image row stays empty
    % 3.1&% mean beta - cones
    % [-8.6,12.1]% CI beta - cones
    % \\
    % \specialrule{0em}{0ex}{2ex}
    % \multicolumn{1}{c}{\fontsize{7}{7}\selectfont Pairwise differences with CIs} \\
    % \specialrule{0em}{0ex}{2ex}
    %%%%%%%% Title row:  Answer Accuracy %%%%%%
    %\specialrule{0em}{4ex}{5ex}
    %\cline{1-3}
     
    \end{tabular}

    \vspace{-0.6cm}
\end{table}

\bpstart{Response Time.} 
Our two studies resulted in different average response times: 5.5-6\,s in Study~1~(\autoref{tab:study01_measurements})
and around 2–2.5\,s in Study~2~(\autoref{tab:study02_measurements}). In Study~1, participants had to coordinate the reading activity and the navigation tasks, which likely resulted in the longer response times. 
The mean difference results in Study~2 show weak evidence that response time with the \slowWalkIcon\,km/h condition was longer than the \fastWalkIcon\,km/h and \slowJogIcon\,km/h conditions, as well as while \standIcon\ standing. The results also show some trends indicating that response time may be shorter with the \slowJogIcon\,km/h condition compared to \sitIcon\ sitting, and while \sitIcon\ sitting compared to the \slowWalkIcon\,km/h condition. 
\revision{The analysis of pairwise differences in Study~1 showed no evidence of a difference between the conditions (see table in the appendix).}

\bpstart{Reading Accuracy.} 
Participants' answers about the fitness data with the highest progress were on average 94.8\% accurate with the three trajectories (Study~1, \autoref{tab:study01_measurements}), 94.7\% with the three walking speeds, and 97\% with the sedentary conditions (Study~2, \autoref{tab:study02_measurements}). Participants also read the percentages on average with a correctness of 72.2\% in Study~1, 78.2\% with the three walking speeds, and 82.2\% with the sedentary conditions. \revision{The analysis of pairwise differences in Study~1 showed no evidence of a difference between the conditions (see table in the appendix).} In general, participants' answers in Study~2 were more accurate with the \standIcon\ standing condition compared to the three walking speeds, in particular the \slowJogIcon\,km/h condition. Interestingly, the results show trends between the sedentary conditions, indicating that reading accuracy might be higher while \standIcon\ standing than while \sitIcon\ sitting. 
For the proportion estimation answers, pairwise differences show weak evidence that participants were less accurate with the \slowJogIcon\,km/h condition in comparison to \standIcon\ standing. In addition, the results show trends that the proportion reading was more accurate with the \slowWalkIcon\,km/h walking speed and while \sitIcon\ sitting compared to the \slowJogIcon\,km/h condition~(\autoref{tab:study02_measurements}). %\newline

\vspace{0.1cm}
\noindent Ultimately, our reading performance measurements show that participants made a trade-off between reading accurately and responding in a reasonable amount of time---that does not overly distract them from the primary task---while walking at different speeds.
%%%%%%%%%%%%%%%%%

%% Measurements Study~2
\setlength{\picturewidth}{.45\linewidth}
\setlength{\pictureheight}{1.65cm}
\setlength{\meanfigwidth}{.1\linewidth}
\setlength{\meanfigheight}{1.6cm}
\setlength{\difffigwidth}{.2\linewidth}
\setlength{\difffigheight}{2.65cm}
\begin{table}[ht!]
\renewcommand{\arraystretch}{1.3}
\centering
    \caption{Means (top) and pairwise differences (bottom) of the reading performance measurements for Study~2: response time (Row~1), fitness data comparison accuracy (Row~2), proportion estimation accuracy (Row~3). Error bars represent 95\% Bootstrap confidence intervals (CIs) in black, adjusted for ten pairwise comparisons with Bonferroni correction (in red).\\ \small \standIcon\ Standing, \sitIcon\ Sitting, \slowWalkIcon km/h, \fastWalkIcon km/h, \slowJogIcon km/h.
    \vspace{-0.2cm}\label{tab:study02_measurements}}
    \fontsize{5.1}{5.29}\selectfont
    \begin{tabular}{@{}p{1.8\picturewidth}@{}%for the 1st image
                    @{}r@{~}%for writing the average
                    @{}p{1cm}@{}%for writing the CI
                    }
     %%%%%%%% Title row: Response Time %%%%%%
    \specialrule{0em}{0ex}{0ex}
    \multicolumn{1}{c}{\small \textbf{Response Time -- in seconds: Study~2}} \\
    \specialrule{0em}{0ex}{2ex}
    \multirow{1}{*}{\includegraphics[height=\meanfigheight]{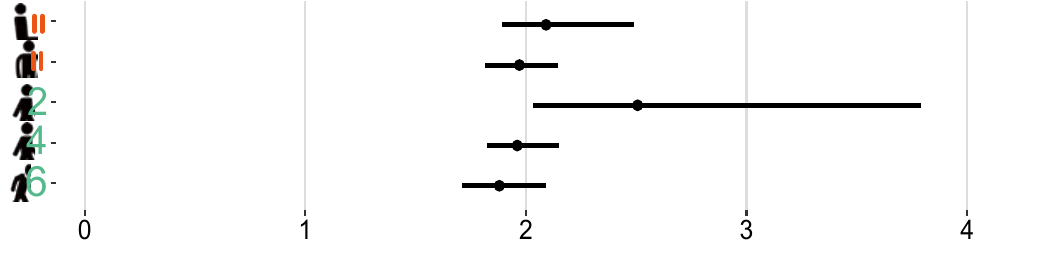}}
     & 2.09 & [1.89, 2.49] \\
     & 1.97 & [1.81, 2.14] \\
     & 2.51 & [2.03,3.79] \\
     & 1.96 & [1.82, 2.15] \\
     & 1.88 & [1.71, 2.09] \\
    \specialrule{0em}{0ex}{4ex}
    \multicolumn{1}{c}{\fontsize{7}{7}\selectfont Means with CIs}\\
    \specialrule{0em}{0ex}{2ex}
    \multirow{1}{*}{\includegraphics[height=\difffigheight]{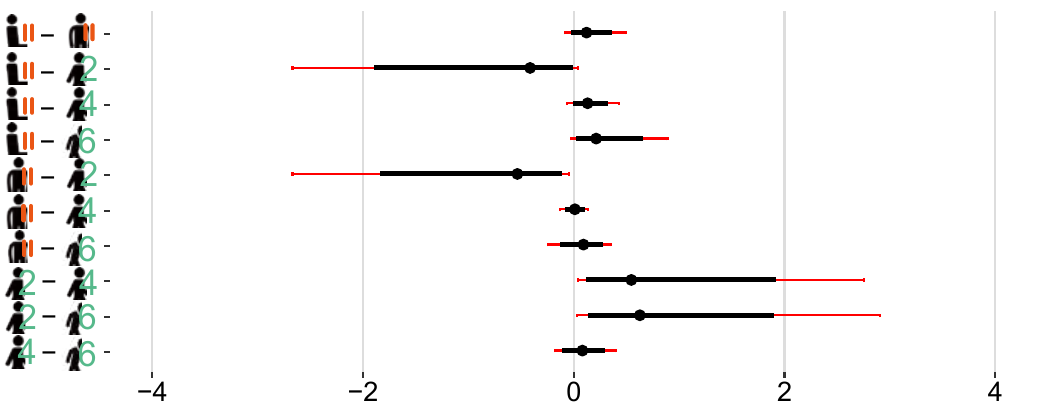}} 
    & 0.12 & [-0.08, 0.49] \\ 
    & -0.42 & [-2.67, 0.04] \\
    & 0.13 & [-0.07, 0.43]  \\
    & 0.21 & [-0.03, 0.89]  \\
    & -0.54 & [-2.67,-0.05] \\
    & 0.01 & [-0.13, 0.14] \\
    & 0.09 & [-0.24, 0.35] \\
    & 0.55 & [0.04, 2.75] \\
    & 0.63 & [0.03, 2.90] \\
    & 0.08 & [-0.18, 0.40] \\
    \specialrule{0em}{0ex}{5ex}
    \multicolumn{1}{c}{\fontsize{7}{7}\selectfont Pairwise differences with CIs} \\ 
    \specialrule{0em}{0ex}{2ex}
    %%%%%%%% Title row: Response time %%%%%%
    \cline{1-3}
    \specialrule{0em}{0ex}{2ex}
    %%%%%%%% Title row: Fitness Data Accuracy %%%%%%
    \multicolumn{1}{c}{\small \textbf{Fitness Data Comparison Accuracy -- in percentage: Study~2}} \\
    \specialrule{0em}{0ex}{2ex}
    \multirow{1}{*}{\includegraphics[height=\meanfigheight]{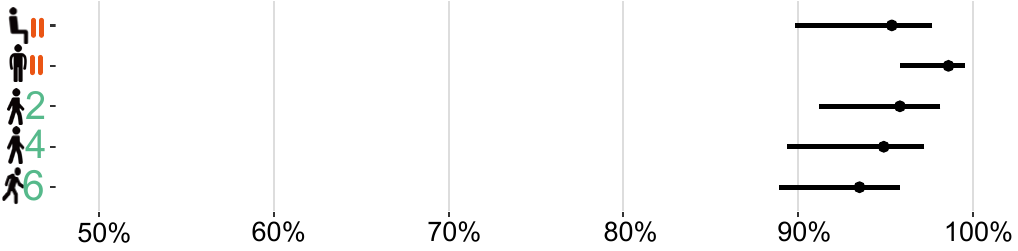}}
     & 95.4  & [89.8, 97.7]\\
     & 98.6  & [95.8, 99.5]\\  
     & 95.8  & [91.2, 98.1]\\
     & 94.9  & [89.4, 97.2]\\
     & 93.5  & [88.9, 95.8]\\
    \specialrule{0em}{0ex}{6ex}
    \multicolumn{1}{c}{\fontsize{7}{7}\selectfont Means with CIs}\\
    \specialrule{0em}{0ex}{2ex}
    \multirow{1}{*}{\includegraphics[height=\difffigheight]{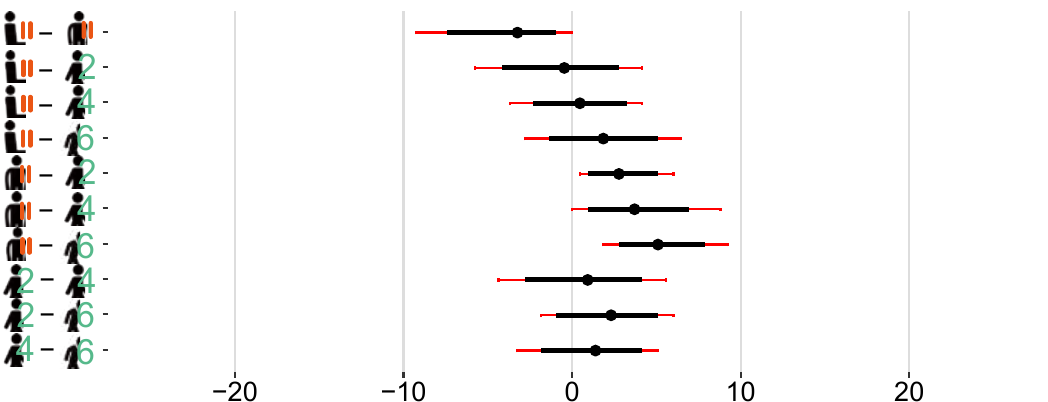}} 
    &-3.2  & [-9.3, 0.0]\\
    &-0.5  & [-5.8, 4.2] \\
    &0.5   & [-3.7, 4.2]\\
    &1.9   & [-2.8, 6.5] \\
    &2.8   & [0.5, 6.0]\\
    &3.7   & [0.0, 8.8]\\
    &5.1   & [1.9, 9.3]\\
    &0.9   & [-4.4, 5.6]\\
    &2.3   & [-1.9, 6.0] \\
    &1.4   & [-3.2, 5.1]\\
    \specialrule{0em}{0ex}{5ex}
    \multicolumn{1}{c}{\fontsize{7}{7}\selectfont Pairwise differences with CIs} \\ 
    \specialrule{0em}{0ex}{2ex}
    %%%%%%%% Title row: Fitness Data Accuracy %%%%%%
    \cline{1-3}
    \specialrule{0em}{0ex}{2ex}
    %%%%%%%% Title row: Answer Proportion Estimation %%%%%%
    \multicolumn{1}{c}{\small \textbf{Proportion Estimation Accuracy -- in percentage: Study~2}} \\
    \specialrule{0em}{0ex}{4ex}
    \multirow{1}{*}{\includegraphics[height=\meanfigheight]{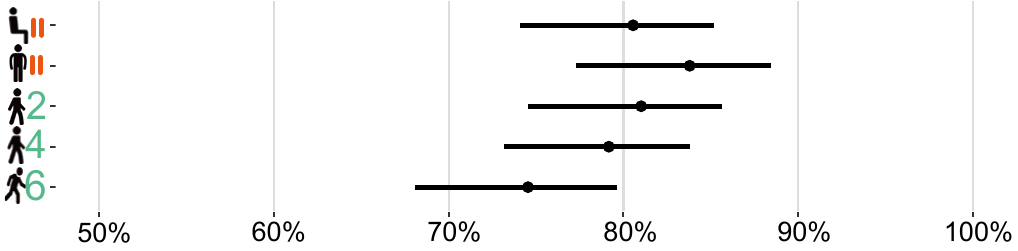}}
     & 80.6  & [74.1, 85.2]\\
     & 83.8  & [77.3, 88.4]\\
     & 81.0  & [74.5, 85.6]\\
     & 79.2  & [73.1, 83.8]\\
     & 74.5  & [68.1, 79.6]\\
    \specialrule{0em}{0ex}{6ex}
    \multicolumn{1}{c}{\fontsize{7}{7}\selectfont Means with CIs}\\
    \specialrule{0em}{0ex}{2ex}
    \multirow{1}{*}{\includegraphics[height=\difffigheight]{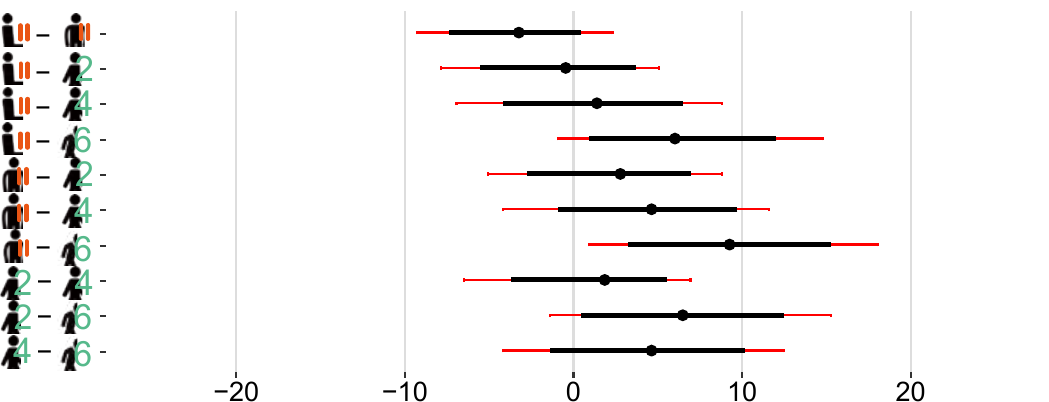}} 
    & -3.2 & [-9.3, 2.3] \\
    & -0.5 & [-7.9, 5.1] \\
    & 1.4  & [-6.9, 8.8] \\
    & 6.0  & [0.9, 14.8] \\
    & 2.8  & [-5.1, 8.8] \\
    & 4.6  & [-4.2, 11.6] \\
    & 9.3  & [0.9, 18.1] \\
    & 1.9  & [-6.5, 6.9] \\
    & 6.5  & [-1.4, 15.3] \\
    & 4.6  & [-4.2, 12.5] \\
    \specialrule{0em}{0ex}{5ex}
    \multicolumn{1}{c}{\fontsize{7}{7}\selectfont Pairwise differences with CIs} \\ 
    \specialrule{0em}{0ex}{4ex}
    %%%%%%%% Title row: Data Proportion Estiamtion Accuracy %%%%%%
    \end{tabular}

    \vspace{-1cm}
\end{table}

\subsection{RQ2: The Influence of Motion on Task Load} %Subjective Measurements}
%\revision{In Tab.~\ref{tab:study1_questionnaires} and Tab.~\ref{tab:study2_questionnaires}} 
We report on participants' subjective evaluation \revision{(see supplemental material for more details)}.  %of the two studies. %\revision{For more details see Table\autoref{tab:study1_questionnaires} and \autoref{tab:study2_questionnaires}.}

\bpstart{Confidence in Provided Answers.} Participants were fairly confident with their responses while walking the three trajectories (median:~5/7). Likewise, their confidence was also high with all the conditions of Study~2 (median:~6/7) except for the \slowJogIcon\,km/h condition, which was rated with a median of 5/7.

\bpstart{Task Difficulty.} Participants found the reading task with the \circularIcon~circular trajectory easier (median:~3/7) than the two other conditions (median:~3.5 and~4/7). This result is echoed in some of their comments: \emph{``It was easier to read data while walking in long continuous circles than short straight lines.''} Another participant mentioned: \emph{``There were less obstacles and I knew I had to walk in the same circular trajectory [...] it took less effort to read the data and compare and answer.''} In Study~2, participants found the reading task easy (median:~2/7) but slightly more challenging with the \slowJogIcon\ km/h condition (median:~3/7). The task difficulty was rated as very easy while \sitIcon\ sitting and~\standIcon\ standing.
    
\bpstart{Mental and Physical Load.}
In Study~1, participants found that the task was more mentally (median:~3.5 and 4/7) than physically demanding with a median of 1/7 for the \lineIcon\ line trajectory and a median of 2/7 for the \circularIcon\ circular and the \infinityIcon\ infinity trajectories. %, particularly with the \infinityIcon\ infinity trajectory. 
One participant mentioned this when reporting about the \infinityIcon\ infinity trajectory: \emph{``The task is not really more challenging or difficult but a little more mentally demanding [...] I'm also thinking about where I will be going at the same time.''} This confirms again that synchronizing the reading and navigation tasks in the first study required more attention from participants. In contrast, in Study~2, participants did not find the task mentally demanding (median:~2/7). However,  the \slowJogIcon~km/h condition was rated more physically demanding (median:~3/7) than the other conditions, with a median of 2/7 for the \fastWalkIcon\,km/h condition and a median of 1/7 for the other conditions. %for the \slowWalkIcon\ km/h and the sedentary conditions.

\subsection{Detail Level of the shown Information}
 In Study 1, some participants mentioned that reading data with this level of precision did not seem a natural task to them: \emph{``in real life, I don't even think that I'll read the \emph{real} value, only intuition about if I'm near the goal or not.''}  In Study 2, we, therefore, gathered additional information from the 12 participants who reported owning a smartwatch or a fitness band. With a multiple-choice question, we asked them to select all the possible options corresponding to the detail level of information they are interested in reading from their devices. We obtained 27 answers, including 17 answers in favor of overviews: 9 ``getting an overview of the displayed data,'' 6~``looking for an approximate value'' (e.\,g., achieved more than half of a target goal), and 2 ``checking data trends.'' The remaining 10 answers were about checking details: 6 ``reading detailed values'' and 4 ``comparing multiple values.'' Even though the majority of the answers were about checking overviews, micro visualizations on smartwatches have their place to convey information at a holistic yet detailed scale---that can be read on demand if necessary. As such, it will be important to tease out how different kinds of reading, search, and identification tasks influence the correctness of reading visualizations while walking. In general, we found that participants made fewer mistakes when reading percentages at the quarters of the radial progress chart (i.\,e., 50\%, 75\%, and 25\%). In contrast, the further the percentages were from the radial progress chart's quarters, the more error-prone participants were \revision{(see supplemental material)}. %Our results show that most errors were made when reading the proportion with 65\%. 
 Accordingly, some participants suggested adding tick marks on the radial bar chart to facilitate the percentage estimation.
\section{Discussion and Conclusion}
Reading micro visualizations on smartwatches can be a challenging task, especially when reading details (e.\,g., proportion estimation of micro visualizations). We found that performing the reading task in motion created additional challenges, such as increased mental load when coordinated with a navigation task or the instability of the watch due to body oscillation from high-speed motion. Our results showed that walking negatively impacted reading performance in general. However, people could read data relatively accurately while walking different trajectories and speeds of up to 6\,km/h. This is good news, as it means that fitness data on smartwatches, similar to ours, can be read effectively on the go. \revision{Yet, now that we know that walking has an impact, several other research questions have opened up. First, it would be interesting to compare our results to other types of activities with continuous long-term movement. These activities could be similar but with more demanding primary tasks that require watching the ground closely or navigating in an unknown terrain such as cross-country running or downhill biking. In addition, more work is needed to test how changes in the representation can improve reading accuracy; for example, through changes in contrast, visualization size, or visualization type.  Solving these questions has concrete implementation opportunities for watch faces that change based on the types of detected movement a wearer engages in or the environment in which they conduct their activity. We encourage future research on effective micro visualizations for sports activities.}

\vspace{-0.05cm}
\acknowledgments{
This project was funded by the DFG grant DFG-ER272-14, DFG under Germany's Excellence Strategy (EXC 2075--390740016),  ANR-18-CE92-0059-01, and ANR-19-CE33-001. Tanja Blascheck is funded by the ESF and the MWK Baden-Württemberg.}

\bibliographystyle{abbrv-doi-narrow}

\bibliography{template}

\newpage
% \appendix
%\input{content/appendices}

\end{document}